\documentclass{ws-ijmpcs}

\newcommand{\Aphys}{A_{\scriptsize\mbox{phys}}}
\newcommand{\Apure}{A_{\scriptsize\mbox{pure}}}

 \newcommand{\beq}{\begin{eqnarray}}
  \newcommand{\eeq}{\end{eqnarray}}

\begin{document}

\markboth{Y. Hatta}
{Nucleon spin decomposition at twist three}

%
\catchline{}{}{}{}{}
%

\title{Nucleon Spin Decomposition at Twist Three}

\author{YOSHITAKA HATTA}

\address{Yukawa Institute for Theoretical Physics, Kyoto University, Kyoto, 606-8502 Japan\\
hatta@yukawa.kyoto-u.ac.jp}

\maketitle


\begin{abstract}
I review the recent progress in understanding the complete gauge invariant decomposition of the nucleon spin with particular emphasis on its twist structure.

\end{abstract}

\ccode{PACS numbers: 12.38.-t, 13.88.+e}

  \section{Introduction}

In recent years there has been renewed interest in the proper  decomposition of the nucleon spin into the quark and gluon degrees of freedom (see Ref.~\refcite{Lorce:2012rr} and references therein). The problem dates back to the classic paper\cite{Jaffe:1989jz} by Jaffe and Manohar in 1990  written shortly after the EMC discovery of  the `proton spin crisis',\cite{Ashman:1987hv} and has remained elusive since then. The recent surge of interest was triggered by a controversial paper by Chen \emph{et al.}\cite{Chen:2008ag} Their original proposal seemed somewhat cryptic, and the connection to observables in high energy experiments was unclear. However,  it can be reinterpreted and revamped into a well--defined framework of spin decomposition consistent with perturbative QCD  in which one can speak of familiar technical language like `twist'. In this short note I summarize the present understanding of the problem from my perspective. The details can be found in Refs.~\refcite{Hatta:2011zs,Hatta:2011ku,Hatta:2012cs,Hatta:2012jm}.

\section{Twist--two decomposition a.k.a. Ji decomposition}

Let me begin with the Ji sum rule\cite{Ji:1996ek}
\beq
J^q = \frac{1}{2}\int_{-1}^1 dx\, x \left(H_q(x) +E_q(x)\right)\,, \qquad
J^g = \frac{1}{4} \int_{-1}^1 dx \left(H_g(x)+E_g(x)\right)\,, \label{11}
\eeq
 which relates the quark/gluon contribution to the nucleon spin ($J^q+J^g=\frac{1}{2}$) to the quark/gluon generalized parton distribution (GPD). I call this twist--two decomposition because the relevant GPDs, $H_{q,g}$ and $E_{q,g}$, are twist--two. The decomposition is based on the (improved) energy momentum tensor of QCD, and as such, all the operators involved are local and gauge invariant. Their matrix elements (i.e., GPDs) are measurable experimentally from deeply--virtual Compton scattering (DVCS) and also by lattice QCD simulations. It is thus a perfectly well--defined decomposition of the nucleon spin based on a firm theoretical background.

 However, this is not the end of the story. There are several important questions which come to mind. 
 \begin{itemize}
 \item What happened to $\Delta G$, the gluon polarization? There has been so much effort, both experimentally and theoretically, to extract this quantity in the QCD spin physics community. But it  doesn't even exist in the above decomposition.

 \item Can one interpret the integrand in (\ref{11}) as a sort of `angular momentum density', with $x$ being the longitudinal momentum fraction of quarks and gluons?

 \item $J^q$ (but not $J^g$) can be further decomposed, gauge invariantly, into the  helicity and the orbital angular momentum (OAM) parts: $J^q = \frac{1}{2}\Delta \Sigma + L^q$. However, this \emph{kinetic} OAM $L^q$ does not satisfy the canonical commutation relation because it involves the covariant derivative $L^q \sim \vec{x} \times \vec{D}$, and $[D^i,D^j]=igF^{ij}\neq 0$. There is an opinion\cite{Ji:2010zza} that each element of a given decomposition need not (and actually cannot, if one considers the quantum evolution) satisfy the commutation relation. But it would be nice to have the \emph{canonical} OAM which satisfies the commutation relation at least at the tree level.

 \item Is the decomposition (\ref{11}) relevant to the longitudinal polarization, or the transverse polarization, or both? Is it frame--independent?
 \end{itemize}

Actually, these questions remain unanswered within the twist--two decomposition.  In order to answer them, one has to go to twist--\emph{three}.

\section{Complete gauge invariant decomposition}

Here is the complete decomposition originally proposed by Chen \emph{et al.}\cite{Chen:2008ag,Chen:2009mr} and written in this `covariant' form by Wakamatsu\cite{Wakamatsu:2010cb}
\beq
 M_{\scriptsize\mbox{quark-spin}}^{\mu\nu\lambda} &=& -\frac{1}{2}\epsilon^{\mu\nu\lambda\sigma}\bar{\psi}
\gamma_5 \gamma_\sigma  \psi\,, \label{23} \\
 M_{\scriptsize\mbox{quark-orbit}}^{\mu\nu\lambda}&=&\bar{\psi}\gamma^\mu (x^\nu iD_{\scriptsize\mbox{pure}}^\lambda
-x^\lambda iD_{\scriptsize\mbox{pure}}^\nu )\psi\,,  \label{24} \\
 M_{\scriptsize\mbox{gluon-spin}}^{\mu\nu\lambda}&=&  F_a^{\mu\lambda}A_{\scriptsize\mbox{phys}}^{\nu a} -
F_a^{\mu\nu}A_{\scriptsize\mbox{phys}}^{\lambda a}  \,,  \label{glu}  \label{25} \\
 M_{\scriptsize\mbox{gluon-orbit}}^{\mu\nu\lambda}&=&  -F_a^{\mu\alpha}\bigl(x^\nu (D_{\scriptsize\mbox{pure}}^\lambda A_\alpha^{\scriptsize\mbox{phys}})_a
-x^\lambda (D^\nu_{\scriptsize\mbox{pure}}A_\alpha^{\scriptsize\mbox{phys}})_a \bigr)\,. \label{26}
\eeq
$A^\mu_{\scriptsize\mbox{phys}}$ is the `physical part' of the gauge field which transforms homogeneously under gauge rotations $\Aphys \to U\Aphys U^\dagger$. The difference $\Apure = A-\Aphys$ is pure gauge (i.e., it is a gauge rotation of the vacuum configuration), and appears in the modified covariant derivative $D_{\scriptsize\mbox{pure}}^\mu \equiv \partial^\mu + igA^\mu_{\scriptsize\mbox{pure}} = D^\mu -ig\Aphys^\mu$. There has been a lot of controversy as to  what exactly $\Aphys$ is.\cite{lorce,wakamatsu} Also, the seeming `covariance' has to be taken with great care.
 One can avoid these subtleties by working in the infinite momentum frame which is the only frame where the partonic picture makes sense and connections to high energy experiments can be established. {\it My} choice is\cite{Hatta:2011zs}
 \beq
A^{\mu }_{\scriptsize\mbox{phys}}(x) = -\int dy^- {\mathcal K}(y^- -x^-) {\mathcal W}_{xy}
F^{+\mu}(y^-,\vec{x}){\mathcal W}_{yx}\,, \label{hatta}
\eeq
where ${\mathcal W}_{xy}$ is the Wilson line from $y^-$ to $x^-$, and
 ${\mathcal K}(y^-)$ is either $\frac{1}{2}\epsilon(y^-)$, $\theta(y^-)$ or $-\theta(-y^-)$. (\ref{hatta}) obviously transforms homogeneously under gauge rotations, and it can be shown\cite{Hatta:2011zs} that the difference $A-\Aphys$ is pure gauge. 
With this definition, I can write
\beq
\frac{1}{2} = \frac{1}{2}\Delta \Sigma + \Delta G + L^q_{can} + L^g_{can}\,, \label{jj}
\eeq
 where each term on the right--hand--side is the appropriate matrix element of the corresponding operator in (\ref{23})--(\ref{26}). Note that this is a complete, gauge invariant decomposition which features $\Delta G$ as the gluon helicity part. It is actually the  gauge invariant completion of Jaffe--Manohar.\cite{Jaffe:1989jz} The quark OAM $L^{q}_{can}$ is different from $L^q$, and the gluon OAM $L_{can}^g$ can be defined. They are the \emph{canonical} OAMs which satisfy the commutation relation because of the property $[D^i_{\scriptsize\mbox{pure}}, D^j_{\scriptsize\mbox{pure}}]=igF^{ij}_{\scriptsize\mbox{pure}}=0$. The expression (\ref{jj}) itself quite often appears in the literature and in presentations, but the precise gauge invariant definitions of $L_{can}^{q,g}$ are seldom articulated or glossed over. I wish to stress that, by showing the relation (\ref{jj}) with the experimentally measurable $\Delta G$, one is implicitly accepting the above decomposition (\ref{23})--(\ref{26}) with $\Aphys$ as given by (\ref{hatta}).

\section{Orbital angular momentum}

The difference between the kinetic OAM $L^q$ and the canonical $L^q_{can}$ is called the potential OAM $L_{pot}$.\cite{Wakamatsu:2010qj} Its operator definition is
\beq
 \bar{\psi}\gamma^+ \left(x^i \Aphys^j -x^j \Aphys^i \right)\psi\,. \label{three}
\eeq
This is gauge invariant by itself.
Inserting (\ref{hatta}) and noticing that $F^{+i}=E^i + (\vec{v}\times \vec{B})^i$ is basically the color Lorentz force, one sees that the above operator is physically interpreted as torque experienced by a  quark as it propagates through the nucleon wavefunction.\cite{Burkardt:2012sd} By taking the nonforward matrix element of (\ref{three}), I can eliminate the transverse coordinate $x^i$ and replace it with the  derivative with respect to the momentum transfer $\Delta^i$.  The remaining quark--gluon operator resembles the one familiar  in the collinear twist--three mechanism of single--spin asymmetry (SSA). Guided by this analogy, I define  doubly--unintegrated densities
\beq
&&\int \frac{dy^- dz^-}{(2\pi)^2} e^{\frac{i}{2}(x_1+x_2)\bar{P}^+z^- + i(x_2-x_1)\bar{P}^+y^-} \nonumber \\
 && \qquad \qquad \times \langle P'S'| \bar{\psi}(-z^-/2)
 \gamma^+
{\mathcal W}_{\frac{-z}{2}y}\, gF^{\mu\nu}(y^-) {\mathcal W}_{y\frac{z}{2}} \psi(z^-/2)|PS\rangle \nonumber \\
 &&= \frac{1}{\bar{P}^+}\epsilon^{\mu\nu\rho \sigma}\bar{S}_\rho \bar{P}_\sigma \Psi(x_1,x_2) + \frac{1}{\bar{P}^+} \epsilon^{\mu\nu\rho\sigma}\bar{S}_\rho \Delta_\sigma \Phi_F(x_1,x_2)+ \cdots\,, \label{nonf}
\eeq
where $\Delta = P'-P$. $x_1$ and $x_2-x_1$ are the momentum fractions assigned to the outgoing quark and gluon, respectively. The first term is relevant to SSA (in the transversely polarized case). The second term is  relevant to the longitudinally polarized case, and is related to the potential OAM\cite{Hatta:2011ku}
\beq
L_{pot}= \int dx_1dx_2 {\mathcal P}\frac{\Phi_F(x_1,x_2)}{x_1-x_2}\,. \label{po}
\eeq

 Speaking of the collinear twist--three approach to SSA, I recall that SSA has an alternative description in terms of the transverse momentum dependent distribution (TMD). This motivates me to define the nonforward generalization of TMD
\beq
&& \int \frac{dz^- d^2z_T}{(2\pi)^3} e^{ix\bar{P}^+z^- -iq_T \cdot z_T}
 \langle P'S'|\bar{\psi}(-z^-/2,-z_T/2)\gamma^+ {\mathcal W}_{\frac{-z_T}{2}, \frac{z_T}{2}}\psi(z^-/2,z_T/2)|PS\rangle \nonumber \\
&& \quad \sim \frac{i}{\bar{P}^+}\epsilon^{+-ij}\bar{S}^+q_{Ti} \Delta_j \tilde{f}(x,q_T^2,\xi,\Delta_T\cdot q_T)\,, \label{bou}
\eeq
 where the Wilson line is U--shaped along the light--cone direction extending to $z^- = \pm \infty$. The matrix elements like (\ref{bou}) were previously defined and classified in  Ref.~\refcite{Meissner:2009ww} where they were called `generalized parton correlation functions'.
 It can be shown\cite{Hatta:2011ku} that the canonical OAM is given by the following moment of $\tilde{f}$ (called $F_{1,4}$ in Ref.~\refcite{Meissner:2009ww}).
 \beq
L_{can}^q =  \frac{1}{2}\int dx d^2q_T\, q_T^2 \tilde{f}(x,q_T^2)\,. \label{oam}
\eeq
Actually, the equation (\ref{oam}) was first derived by Lorce and Pasquini\cite{Lorce:2011kd} using the Wigner distribution neglecting the Wilson line. Since the Wigner distribution describes the phase space (position and momentum) density of partons, it can be naturally used to define an OAM which is the cross product of the position and the momentum. The question is \emph{which} OAM one gets in this way.   Interestingly, this is determined by the choice of the contour in the Wilson line. As stated above, the U--shaped Wilson line along the light cone leads to the canonical OAM. However, the straight Wilson line leads instead to the kinetic OAM.\cite{Ji:2012sj,Lorce:2012ce}

\section{Twist analysis}

Now I come to the issue of `twist' (Ref.~\refcite{Hatta:2012cs}, see also Ref.~\refcite{Ji:2012ba}).  The two decompositions discussed so far, the Ji decomposition and the complete decomposition, are related as follows
\beq
&&J^q = \frac{1}{2}\Delta \Sigma + L^q_{can} + L_{pot}\,,  \label{q}\\
&& J^g + L_{pot} = \Delta G + L_{can}^g\,. \label{g}
\eeq
Remarkably, these relations can be understood at the density level.\cite{Hatta:2012cs} Actually, it is possible to uniquely (in a certain sense) define the density of the canonical OAM $L_{can}^{q,g} = \int dx L_{can}^{q,g}(x)$. This allows me to analyze the twist structure of the complete decomposition, and in particular, its relevance to  twist--\emph{three} GPDs.

Let me begin with the relation $L^q = L_{can}^q + L_{pot}$. $L^q$ involves the `D--type' correlator $\bar{\psi}\gamma D^i \psi$ and $L_{pot}$ involves the `F--type' correlator $\bar{\psi} \gamma F^{+i}\psi$. It is known that these two types of correlators are related.\cite{Eguchi:2006qz} In terms of the doubly--unintegrated densities defined similarly to the second term of (\ref{nonf})
\beq
F.T.\langle P'S'| \bar{\psi}\gamma^+ F^{+i}\psi |PS\rangle &\sim&  \Phi_F(x_1,x_2)\,, \nonumber \\
F.T.\langle P'S'| \bar{\psi}\gamma^+\gamma_5 F^{+i}\psi |PS\rangle &\sim&  \widetilde{\Phi}_F(x_1,x_2)\,, \nonumber \\
F.T.\langle P'S'| \bar{\psi}\gamma^+ D^i \psi |PS\rangle &\sim& \Phi_D(x_1,x_2)\,, \nonumber \\
F.T.\langle P'S'| \bar{\psi} \gamma^+\gamma_5 D^i \psi |PS\rangle &\sim& \widetilde{\Phi}_D(x_1,x_2)\,,
\eeq
  the relation reads 
  \beq
  \Phi_D(x_1,x_2) = \delta (x_1-x_2) L_{can}^q(x_1)+{\mathcal P} \frac{1}{x_1-x_2} \Phi_F(x_1,x_2) \,, \label{can}
  \eeq
   which is the doubly--unintegrated version of $L^q = L_{can}^q + L_{pot}$ (cf. (\ref{po})).
Eq.~(\ref{can}) naturally defines the canonical OAM density $L_{can}^q = \int dx L_{can}^q(x)$.
 The delta function $\delta(x_1-x_2)$ ensures that, in the quark--gluon--quark system described by the operator $\bar{\psi}D\psi$, the gluon carries zero longitudinal momentum $x_2-x_1=0$. Thus the variable $x$ in $L_{can}^q(x)$ is indeed the longitudinal momentum fraction assigned to the quark, which makes its density interpretation preferable. In contrast, there is an ambiguity in defining a `density of the kinetic OAM' $L^q = \int dx L^q(x)$.  For instance, one can define either $L^q(x) = \int dx' \Phi_D(x,x')$ or $L^q(x)=\int dx' \Phi_D(x+x'/2,x-x'/2)$.

The expression of $L_{can}^q(x)$ is complicated, but owing to the equation of motion it can be written as\cite{Hatta:2012cs}
\beq
L_{can}^q(x)  &=& x(H_q(x)+E_q(x)+G_3(x) ) -  \Delta q(x) \nonumber \\
 && \qquad -\int dx' {\mathcal P}\frac{1}{x-x'} \left(\Phi_F(x,x') +\widetilde{\Phi}_F(x,x')\right)\,,  \label{obtain}
\eeq
 where $G_3(x)$ is one of the twist--\emph{three} GPDs defined as  
 \beq
 F.T.\langle P'S'|\bar{\psi}(-z/2)\gamma^i_\perp \psi(z/2) |PS\rangle = G_3(x) \bar{u}(P'S')\gamma^i_\perp u(PS)+\cdots\,. \label{over}
 \eeq
 Integrating (\ref{over}) over $x$, I get 
 \beq
 \int dx\, x G_3(x) = -L^q\,. \label{mark}
 \eeq
 This identity was first derived in Ref.~\refcite{Penttinen:2000dg}. However, there the authors worked in the parton model where there is no distinction between $L^q$ and $L^q_{can}$.   (\ref{obtain}) and (\ref{mark}) show that, while the integral of $G_3$ is related to the kinetic OAM, $G_3(x)$ itself is rather related to the canonical OAM. 
 
Furthermore, $G_3(x)$ can be eliminated from (\ref{obtain}) due again to the equation of motion. The result is
\beq
L_{can}^q(x)
&=& x\int_x^{\epsilon(x)} \frac{dx'}{x'} (H_q(x')+E_q(x')) -x\int_x^{\epsilon(x)} \frac{dx'}{x'^2} \Delta q(x') \nonumber  \\
&& \qquad  -
x\int_x^{\epsilon(x)} dx_1\int_{-1}^1 dx_2\Phi_F(x_1,x_2) {\mathcal P}\frac{3x_1-x_2}{x_1^2(x_1-x_2)^2}  \nonumber \\
&& \qquad -x\int_x^{\epsilon(x)} dx_1\int_{-1}^1 dx_2\widetilde{\Phi}_F(x_1,x_2)  {\mathcal P}\frac{1}{x_1^2(x_1-x_2)}\,,
 \label{mom}
\eeq
 where $\Delta q$ is the usual polarized quark distribution. 
Eq.~(\ref{mom}) completely reveals the twist structure of $L_{can}^q(x)$. It can be decomposed into the `Wandzura--Wilczek' part which is related to the twist--two GPDs, and the `genuine twist--three' part. Taking the first moment of (\ref{mom}), I get
\beq
L_{can}= J^q  -\frac{1}{2}\Delta \Sigma -L_{pot}\,,
\eeq
which is precisely  (\ref{q}).

Similarly, I can define the canonical gluon OAM density $L_{can}^g(x)$ and analyze its twist structure. Again, the definition is unique in the sense that $x$ is  interpretable as the longitudinal momentum fraction of the outgoing gluon. As in (\ref{obtain}), the density is related to one of the twist--three\emph{ gluon} GPDs. By eliminating the twist--three GPD using the equation of motion, I get the  decomposition of $L_{can}^g(x)$ into the part related to the twist--two gluon GPDs and genuine twist--three, three gluon distributions.  
 Its first moment of course coincides with (\ref{g}).

\section{Transverse spin decomposition}

Actually, in the discussions so far I implicitly assumed that the spin is longitudinally polarized. In the transversely polarized case, the situation is a bit more subtle.  Firstly, one has to use the Pauli--Lubanski vector\cite{Ji:2012vj}
\beq
W^\mu = -\frac{1}{2}\epsilon^\mu_{\ \, \nu\rho\sigma}P^\nu  \int d^3x M^{+\rho\sigma}\,, \label{d3}
 \eeq
 instead of the angular momentum tensor $J^{\mu\nu}=\int M^{+\mu\nu}$ itself. The reason is that the latter cannot give a frame--independent decomposition because rotation and boost do not commute.\cite{Leader:2012md}
The relevant component is
 \beq
 W^i = \epsilon^{ij}\left(P^- \int d^3x M^{++}_{\quad \ j} -P^+\int d^3x M^{+-}_{\ \ \ \ j} \right)\,, 
\eeq
  where
\beq
M^{++j}=x^+ T^{+j}-x^j T^{++}\,, \quad M^{+-j}=x^- T^{+j}-x^jT^{+-}\,.
\eeq 
 Note that $W^i$ involves different components of $T^{\mu\nu}$ with  different twists. Their matrix elements are related by Lorentz symmetry, and this leads to the cancelation of some unwanted, frame--dependent terms.\cite{Ji:2012vj} However, one cannot eliminate the frame--dependence completely. The matrix element of the twist--\emph{four} operator $T^{+-}_{q,g}$ contains a term proportional to the metric tensor\cite{Ji:1996ek}
 \beq
 \langle P'S'| T^{+-}_{q,g}|PS\rangle \sim g^{+-} \bar{C}_{q,g}\bar{u}(P'S')u(PS)\,,
 \eeq
  which has no counterpart in the matrix elements of $T^{++}_{q,g}$ and $T^{+i}_{q,g}$ (because $g^{++}=g^{+i}=0$). Then the question is whether this term does any harm, and unfortunately the answer is yes.  As observed in Ref.~\refcite{Bakker:2004ib}, the \emph{non}forward product of spinors $\bar{u}(P'S')u(PS)$ is {\it not} a Lorentz scalar. It contains a manifestly frame--dependent term
 \beq
\bar{u}(P'S') u(PS) \approx 2M +i\frac{ \bar{P}^3 }{M(\bar{P}^0 +M)}\epsilon^{ij}\Delta_i S_j\,,
\eeq
 in the transversely polarized case (but not in the longitudinally polarized case). The linear term in $\Delta$  modifies  the Ji sum rule as\cite{Hatta:2012jm,Leader:2012ar}
  \beq
 J^{q,g} \to J^{q,g} +  \frac{P^3}{2(P^0+M)}\bar{C}_{q,g}\,, \label{abc}
 \eeq
  keeping the sum $J^q+J^g$ unchanged because $\bar{C}^q+\bar{C}^g=0$.\footnote{There is a mismatch in the coefficient of $\bar{C}$ between Ref.~\refcite{Hatta:2012jm} and Ref.~\refcite{Leader:2012ar}. The difference comes from the meaning of $d^3x$ in (\ref{d3}). The former used the light--front form $d^3x = dx^- d^2x_\perp$, while the latter used the instant form $d^3x = dx_1dx_2dx_3$. }

The extra term in (\ref{abc}) is a nominally  `higher twist' contribution, but it is numerically not suppressed because all the terms in (\ref{abc}) are expected to be order unity when the nucleon is relativistic $P^3 \approx P^0$.

If the Ji decomposition has a problem, then what about the complete decomposition (\ref{23})--(\ref{26})? 
 A careful analysis shows that the best one can achieve in the transversely polarized case is\cite{Hatta:2012jm}
  \beq
 \frac{1}{2} =\frac{1}{2}\Delta \Sigma + \Delta G + L_{can}\,,
 \eeq
 where $\Delta \Sigma$ and $\Delta G$ are numerically the same as in the longitudinally polarized case and given by the matrix elements of  (\ref{23}) and (\ref{25}). However, the canonical OAM $L_{can}$ cannot be separated into the quark and gluon contributions. Trying to do so will result in frame--dependent terms similar to that encountered in (\ref{abc}).  \\

\vspace{3mm} 
To conclude, I note that all the four questions that I posed in Section 2 have been answered. In the longitudinally polarized case, the  complete gauge invariant decomposition of the nucleon spin---the twist--{\it three} decomposition---is now available even at the  level of the density in $x$.

\section*{Acknowledgments}

I would like to thank the organizers of ``QCD evolution workshop" at Jefferson Lab for a kind invitation to present this work.

\end{document}